\documentclass[useAMS,usenatbib,usenatbib]{mn2e}    
\usepackage{amssymb}  
\usepackage{graphicx}
\def\be{\begin{equation}} 
\def\en{\end{equation}}

\begin{document}

\title{Galaxy cluster  mergers} 
\author[S. Planelles  \& V. Quilis] 
{Susana  Planelles \&  
Vicent  Quilis\thanks{e-mail: vicent.quilis@uv.es} \\
Departament   d'Astronomia    i   Astrof\'{\i}sica,   Universitat   de
Val\`encia,   46100   -    Burjassot   (Valencia),   Spain}  

\date{Received date / Accepted date}

\maketitle
 
\begin{abstract}

We present the  results of an Eulerian adaptive  mesh refinement (AMR)
hydrodynamical and N-body simulation in a $\Lambda$CDM cosmology.  The
simulation  incorporates  common  cooling  and heating  processes  for
primordial gas.  A specific halo  finder has been designed and applied
in order to extract a sample of galaxy clusters directly obtained from
the simulation  without considering any resimulating  scheme.  We have
studied the evolutionary history  of the cluster halos, and classified
them into  three categories depending  on the merger events  they have
undergone: major  mergers, minor  mergers, and relaxed  clusters.  The
main properties of each one of these classes and the differences among
them are discussed. The  collisions among galaxy clusters are produced
naturally by  the non-linear  evolution in the  simulated cosmological
volume, no controlled collisions have been considered.  We pay special
attention to discuss the role of merger events as a source of feedback
and reheating, and their effects on the existence of cool cores in 
galaxy clusters, as well as in the scaling relations. 

\end{abstract}
 

\section{Introduction}

Galaxy  clusters  are  crucial  pieces  in our  understanding  of  the
Universe.   They  represent   the  transition  between  two  different
regimes.   The first one  involves the  evolution of  perturbations on
cosmological scales where the physics of the problem is only driven by
gravity on the dark matter component.  The second one, is the scenario
of galactic scales,  where the action of gravity  is combined with the
effects  of a  complex  gas  dynamics and  all  sort of  astrophysical
phenomena (cooling, star formation, feedback, ...).

Precisely,  due to  its nature  linking different  scales,  the galaxy
clusters provide us with a powerful tool to constrain the cosmological
parameters,  and at  the same  time, they  help us  to  understand the
physical  properties  of  the   intra-cluster  medium  (ICM)  and  its
interplay with galaxy formation  and evolution.  In this sense, galaxy
clusters  are  extremely   valuable  laboratories  where  explore  the
connection between cosmological scales and the formation and evolution
of galaxies.

The   simplest   model   explaining   the  properties   of   the   ICM
~\citep{kaiser86} assumes  that gravity is the  only responsible force
determining  the evolution  of the  ICM.   In this  scenario, the  gas
collapses  into the dark  matter potential  wells and  then, accretion
shocks  form moving outwards  in the  cluster and  heating up  the gas
until the virial temperature of the halo ~\citep{quilis98}. As gravity
acts on all scales, this model has been known as self-similar.

This model gives  us precise predictions for the  properties of galaxy
clusters:   X-ray   luminosity,   temperature,   entropy,   and   mass
~\citep{brynor98}.   However, the  scaling relations  produced  by the
self-similar model do not  match the observational results completely.
More  specifically,  the  slope  of the  X-ray  luminosity-temperature
relation is steeper  than predicted ~\citep{marke98, arev99, ospon04},
the measured  gas entropy in poor  clusters and groups  is higher than
expected  ~\citep{ponman03}, and  it  has been  observed a  decreasing
trend    of    the   gas    mass    fraction    in   poorer    systems
~\citep{balogh01,lin03,sanderson03}.    In  addition,   there   is  an
important scatter  in the relations ~\citep{fabian94}  partly, but not
totally, connected with the effect of the different environments where
clusters live.

Those   discrepancies   between  the   self-similar   model  and   the
observations  have motivated  the  idea that  some important  physics,
basically  related with  the  baryonic component,  is  missing in  the
model.

Some  of  these  non-gravitational processes  have  been  included  in
simulations  trying  to  solve  the  similarity  breaking:  preheating
~\citep{navarro95,bialek01,borgani02},     and    radiative    cooling
~\citep{pearce00,muan01,dave02,molt04,krav05}.    More   sophisticated
approaches coupling the feedback  with cooling and star formation have
been      carried       out      by      \citet{kay03,torna03,valda03,
borgani04,etto04,kay04,kay07} among others.

Merger events  can also be an  important source of  feedback in galaxy
clusters. They  can produce shocks and compression waves in  the halos
which  eventually can  release part of the  energy associated with the
collision as thermal energy in the final system ~\citep{maccar07}.  It
is likely that  turbulence and mixing could play  an important role in
how this  energy is mixed  and released in  the ICM of the  final halo
after the merger.   As it is well known, some of  the results of these
simulations  could  depend  on  the  ability  of  different  numerical
techniques to describe shock  waves, strong gradients, turbulence, and
mixing, which can  be very  different.   Although is  still a
matter of  debate, it  has been  shown, at  least for  some idealised
tests, that  the comparison between grid  codes and SPH  codes -- when
numerical resolution is similar -- can give substantial differences in
the results  ~\citep{frenk99,agertz07}.  It seems  reasonable to think
that  these  inherent   numerical  differences  could  translate  into
relevant  differences  when  they  are  applied to  more  complex  and
realistic  scenarios  like   galaxy  clusters.  This  situation  makes
interesting,  and complementary, to  pursue   the number  of studies
using the different numerical strategies available.

In this paper,  we want to investigate the role  of the galaxy cluster
mergers as  a source of feedback  and reheating in  a complete general
cosmological framework. The galaxy clusters form and evolve due to the
non-linear  evolution of primordial  perturbations, and  therefore, no
special  symmetry  or  idealised  clusters are  considered.   In  this
scenario, the  merger events naturally  take place according to the
hierarchical evolution.   Previous works have  extensively studied the
mergers   of    galaxy    clusters   using    controlled    collisions
(e.g.    \citet{ricker2001,poole06,poole07,maccar07,poole08}).     The
approach  adopted  in  the   present  paper  could  be  considered  as
complementary  to the studies  using controlled  mergers. It  is clear
that our  approach has some important weaknesses, when it  is compared
with   controlled  mergers,   like   the  worst   resolution  or   the
impossibility to control  the  different parameters  involved in  the
problem.  However,  it  gives  a  description  of  the  problem  in  a
cosmological  context, without symmetries,  including the  presence of
substructures  and taking into  account the  effects of  the different
environments.

In  order to  study  the role  of  mergers,  fulfilling all  the
previous requirements, we have carried  out a simulation of a moderate
size box of side length $100 \,h^{-1}\,Mpc$. We have performed it with
an  Eulerian AMR cosmological  code including  the usual  processes of
cooling and  heating for primordial  gas, and a  phenomenological star
formation treatment.  We have identified and
followed the  evolution of the  different galaxy cluster  halos.  Once
the evolutionary  history of  the halos is  known, we  have classified
them into three broad categories 
depending on the features of the merger
events  in which  they  have  been involved.   We  will discuss  their
effects on cluster  properties. These mergers are
the  ones naturally  happening in  the
building up of the galaxy clusters.

The paper is organized as follows. In Section 2, we present the
details  of  the simulation  and  describe  the  halo finder  used  to
identify  the galaxy  cluster halos.   In  Section 3,  we analyse  the
results of the simulation describing the main properties of the galaxy
cluster sample, and the effects of mergers. Finally,  in Section 4, we
summarise and discuss our results.

\begin{table*}
\begin{minipage}{126mm}
\begin{center}
\caption{Cluster sample. Main features  of selected clusters at $z=0$.
Column 2, virial radius in units  of $h^{-1}\,Mpc$; column 3,
total  mass within the virial radius  in units of $10^{14}\,
M_\odot$;  column
4, mass-weighted temperature within the virial radius in $keV$; column
5,  gaseous mass  within  the  virial radius  in  units of  $10^{14}\,
M_\odot$; column 6, average entropy within the virial radius in units 
of $keV cm^2$; 
column 7,  type classification:  
relaxed or with no important
mergers (R), minor  mergers (MI), and major mergers (MA).}
\label{tab1}
\begin{tabular}{ccccccc}
\hline cluster&$r_{vir}$ & $M_{vir}$ & T & $m_{gas}$ & $S$ & type \\
& $(h^{-1}\,Mpc)$ & $(10^{14}\,M_\odot)$ &  $(keV)$   &
  $(10^{14}\,M_\odot)$ & $(keV cm^2)$ & \\ 
\hline 
CL01 & 2.32 & 18.61 & 7.02 & 3.45 & 2601.06 & MA \\
CL02 & 2.22 & 15.83 & 5.99 & 4.89 & 2130.17 & MA \\
CL03 & 1.58 & 5.68 & 3.24 & 1.08 & 919.34 & MA   \\
CL04 & 1.48 & 4.70 & 3.80 & 0.79 & 998.65 & MA   \\
CL05 & 1.39 & 3.93 & 2.29 & 0.70 & 1063.19 & MA  \\
CL06 & 1.07 & 1.86 & 1.26 & 0.37 & 510.01 & MA   \\
CL07 & 1.01 & 1.53 & 1.01 & 0.25 & 846.33 & MA   \\
CL08 & 0.93 & 1.14 & 0.87 & 0.21 & 422.39 & MA   \\
CL09 & 1.51 & 5.19 & 3.09  & 1.03 & 1166.28 & MI \\
CL10 & 1.51 & 5.11 & 2.18 & 0.92 & 1126.22 & MI  \\
CL11 & 1.36 & 3.76 & 2.55 & 0.49 & 862.07 & MI   \\
CL12 & 1.45 & 4.40 & 2.52 & 0.97 & 980.39 & R    \\
CL13 & 1.10 & 1.99 & 1.40 & 0.41 & 409.21 & R    \\
CL14 & 0.99 & 1.39 & 1.18 & 0.29 & 362.36 & R    \\
CL15 & 0.89 & 1.02 & 0.97 & 0.22 & 554.70 & R    \\
CL16 & 0.89 & 1.01 & 0.90 & 0.22 & 386.56 & R    \\
\hline
\end{tabular}
\end{center}
\end{minipage}
\end{table*}

\section{The simulation}

\subsection{Simulation details}

The simulation  described in  this paper has  been performed  with the
cosmological  code  MASCLET \citep{quilis04}.   This  code couples  an
Eulerian  approach  based  on  {\it high-resolution  shock  capturing}
techniques  for describing  the  gaseous component,  with a  multigrid
particle mesh  N-body scheme for evolving  the collisionless component
(dark matter).  Gas and dark matter are coupled by the gravity solver.
Both  schemes  benefit of  using  an  adaptive mesh refinement  (AMR)
strategy, which permits to gain spatial and temporal resolution.

The numerical simulation has  been performed assuming a spatially flat
$\Lambda CDM$  cosmology, with the  following cosmological parameters:
matter  density  parameter,  $\Omega_m=0.25$;  cosmological  constant,
$\Omega_{\Lambda}=\Lambda/{3H_o^2}=0.75$;  baryon  density  parameter,
$\Omega_b=0.045$;  reduced Hubble  constant, $h=H_o/100  km\, s^{-1}\,
Mpc^{-1}=0.73$;  power  spectrum index,  $n_s=1$;  and power  spectrum
normalisation, $\sigma_8=0.8$.

The initial  conditions were  set up at  $z=50$, using a  CDM transfer
function from \citet{EiHu98}, for a  cube of comoving side length $100
\,  h^{-1}\,  Mpc$.  The  computational  domain  was discretized  with
$512^3$ cubical cells.

A first  level of refinement (level 1)  for the AMR scheme  was set up
from the  initial conditions  by selecting regions  satisfying certain
refining criteria,  when evolved  -- until present  time --  using the
Zeldovich  approximation.   The  volumes  selected as  
{\it refinable}  were
covered  by grids  (patches) with  numerical cells  selected  from the
initial conditions. The regions of  the box not eligible to be refined
were degraded  in resolution by  averaging the quantities  obtained on
the initial grid. This procedure creates the coarse grid (level 0) for
the AMR  scheme. These coarse cells  have a volume  eight times larger
than  the first  level  ones.  In  the  same manner,  the dark  matter
component  within the  refined regions  was sampled  with  dark matter
particles eight times lighter than  those used in regions covered only
by the  coarse grid.  During  the evolution, regions on  the different
grids  are  refined  based  on  the local  baryonic  and  dark  matter
densities.    Any    cell   with    a   baryon   mass    larger   than
$5.6\times10^8M_{\odot}$   or   a  dark   matter   mass  larger   than
$2.5\times10^9M_{\odot}$ was  labelled as {\it  refinable}.  The ratio
between the cell sizes for a  given level ($l+1$) and its parent level
($l$)   is,  in   our  AMR   implementation,   $\Delta  x_{l+1}/\Delta
x_{l}=1/2$.  This is a compromise value between the gain in resolution
and  possible numerical instabilities.   This method  produces patches
with a boxy geometry and cubic cells at any level.

The simulation  presented in  this paper has  used a maximum  of seven
levels ($l=7$) of refinement, which gives a peak spatial resolution of
$3\,h^{-1}\,kpc$.   For the  dark  matter we  consider two  particles
species, which correspond  to the particles on the coarse grid and the
particles  within  the  first  level  of  refinement  at  the  initial
conditions.  The  best mass resolution is  $5.75\times 10^8\, h^{-1}\,
M_\odot$, equivalent to distribute $512^3$ particles in the whole box.

Our simulation includes cooling  and heating processes which take into
account Compton and free-free  cooling, UV heating \citep{hama96}, and
atomic and molecular cooling for a primordial gas. In order to compute
the abundances  of each species, we assume  that the gas  is optically
thin  and in ionization  equilibrium, but  not in  thermal equilibrium
\citep{katz96,theuns98}.  The tabulated  cooling rates were taken from
\citet{sudo93} assuming a constant  metallicity 0.3 relative to solar.
The cooling curve was  truncated below temperatures of $10^4\,K$.  The
cooling and heating  were included in the energy  equation (see Eq.(3)
in \citet{quilis04}) as extra source terms.

The star  formation has also  been modelled with  the phenomenological
approach     commonly     used     in     cosmological     simulations
(i.e. \citet{yepes97,springe03}).  A more detailed  description of how
star formation  has been  introduced in MASCLET  code is  presented in
Appendix A. Despite the use of  an AMR code to  perform the simulation
described  in   the  present  paper,  we  have   still  had  numerical
limitations, namely, the number of patches placed at the highest level
of refinement. Although this limitation  has been not crucial for the
description of clusters, it has  translated into a poor star formation
efficiency  as the analysed run  only allowed  star formation  at the
highest  level   of  refinement.    This  apparent  drawback   of  our
simulation  is not dramatic for  the purpose of  the present study
where we focus  in the effect of mergers on  the ICM properties. Thus,
the stellar feedback has turned  out to be very low, and consequently,
it does not alter the pure effect of the mergers.

\begin{figure*}
\centering\includegraphics[width=16cm]{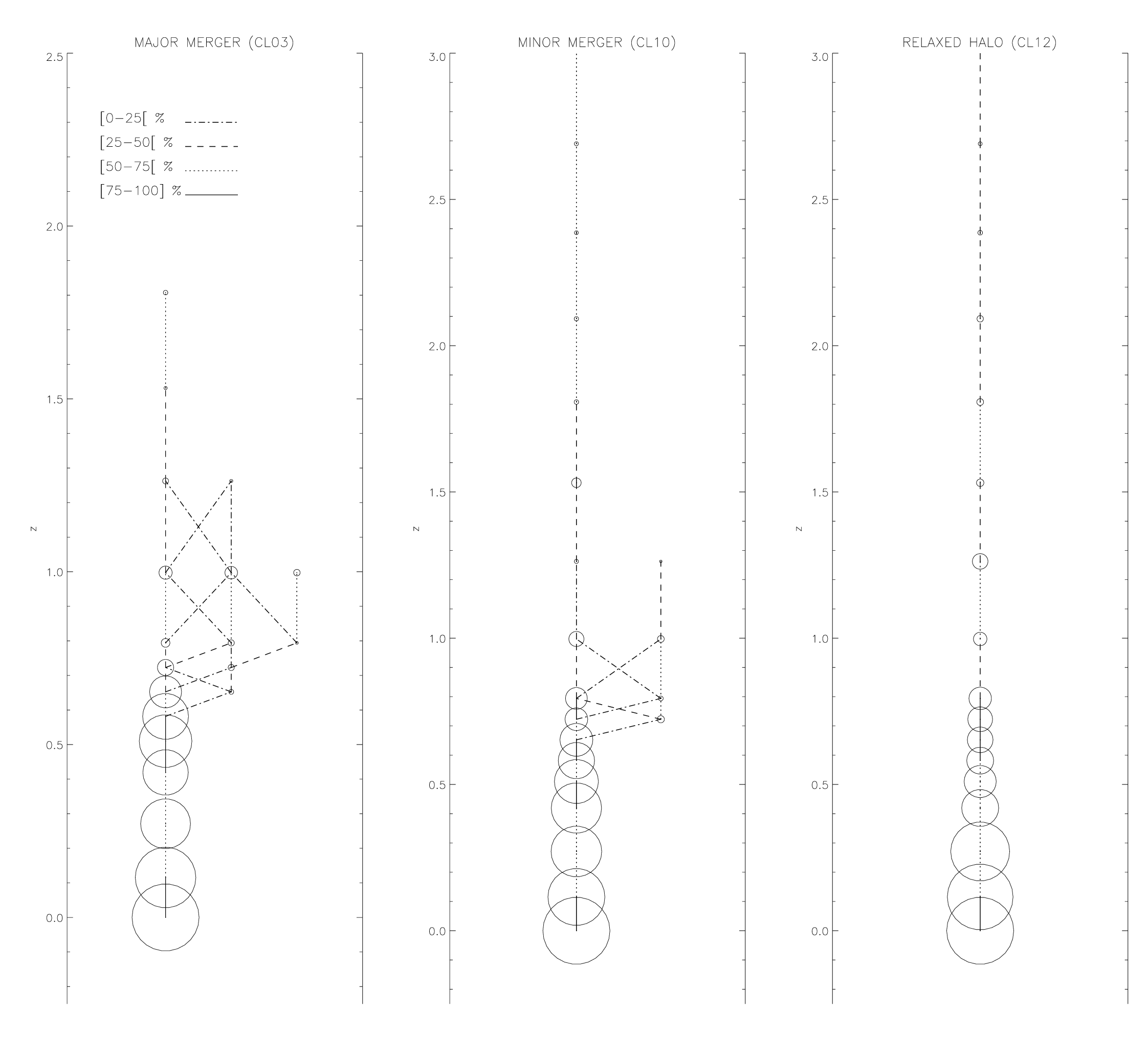}
\caption{Merger tree for one cluster of each category:
left panel shows a   
major merger cluster, central  panel represents a minor 
merger cluster, and 
right panel stands for a relaxed cluster.
Cluster halos are represented by circles whose sizes are normalised 
to the final mass at $z=0$. Lines connecting 
halos at different times indicate the amount of mass transferred from 
the progenitors to their descendants.}
\label{mertree}
\end{figure*}

\subsection{Cluster identification}

A crucial issue  in the analysis of our simulation has  to do with the
cluster identification.  In order to do  so, we have  developed a halo
finding method  
specially suited for  the features of  the cosmological
code MASCLET.

The halo finder  developed for MASCLET code follows  a similar idea to
MHF  halo   finder  \citep{gkg04}.   The  main  aim   is  to  identify
gravitationally bound objects in a N-body simulation.
  
We have  used an identification technique 
based on the  original idea of
the  spherical over-density (SO)  method.  The  basic concept  of this
technique  is to identify  spherical regions  with an  average density
above  a certain  threshold, which  can  be fixed  according to  the
spherical top-hat collapse model. Therefore, we can define 
the viral mass of
a  halo, $M_{vir}$,  as the  mass enclosed  in a  spherical  region of
radius, $r_{vir}$, having  an average  density $\Delta_c$  times the
critical      density      $\rho_{c}(z)=3H(z)^2/8\pi      G$: 
\be
\label{mvir}
M_{vir}(<r_{vir})=\frac{4}{3}\pi r_{vir}^3 \Delta_c \rho_{c}.  
\en

The over-density $\Delta_c$ depends on the adopted cosmological 
model, and can be approximated by the following expression 
\citep{brynor98}:
\begin{equation}
\Delta_c= 18 \pi^2 + 82 x - 39 x^2
\label{Deltac}
\end{equation}
\noindent
where  $x=\Omega(z)-1$  and  $\Omega(z)=[\Omega_m (1+z)^3]/  [\Omega_m
(1+z)^3 + \Omega_\Lambda]$. Typical  values of  $\Delta_c$ are  
between 100  and 500, in particular, for the 
cosmological parameters considered in our simulation,  
$\Delta_c \simeq 373 $.  

The SO method implies  oversimplifications which could lead to somehow
artificial results  which deserve  a careful treatment.   The enforced
spherical symmetry  for halos that, at  least at some moments of their
lives, can  have quite irregular  structures (e.g., ~\citet{white02}),
or  the difficulties to  discriminate among  close density  peaks, are
some of these well known problems.

The  practical implementation  of our  halo finder  has  several steps
designed to improve the performance of the SO method and to get rid of
the possible drawbacks. In a  first step, the algorithm finds halos by
centring  at  density peaks  and  growing  spheres  until the  average
over-density falls  below $\Delta_c$, or there  is a rising  up of the
slope of the  density profile. The second step  takes care of possible
overlaps among the  preliminary halos found in the  first step. In our
method, halos which overlap more than  30\% and less than 80\% in mass
are joined as a single one. Halos sharing more than 80\% of their mass
are considered as a misidentification and one of them drops out of the
list. The third step checks that all particles contained in a halo are
bound. In  order to determine whether  a particle is bound  or not, we
estimate  the  escape  velocity   at  the  position  of  the  particle
\citep{kgk99}. If the velocity of a particle is larger than the escape
velocity, the particle is assumed to be unbound. A final step verifies
that   the  density  profile   is  consistent   with  a   NFW  profile
\citep{nfw97}.

One of  the main  advantages of  our method is  that the  structure of
nested grids  created by  the AMR scheme  already follows  the density
peaks,  and  therefore,  densities   are  already  calculated  by  the
cosmological code.   Other important point, inherent  to the structure
of the  AMR scheme  used, is  that no linking  length is  needed.  The
process of halo finding can be performed, independently, at each level
of  refinement of the  simulation. Then,  in a  natural way,  our halo
finder  can trace  halos-in-halos  and obtain  a  hierarchy of  nested
halos.

The different  progenitors are  identified by following  all particles
belonging to  a given  halo backwards in  time. This procedure  can be
repeated until the first progenitor  of a certain halo is found.  This
method allows us,  not only to know all the  progenitors of each halo,
but the amount of mass received from each one of its ancestors.

\section{Results}

In our simulation,  we have identified more than  three hundred galaxy
clusters   and  groups   spanning  a   range  of   masses   from  $1.0
\times10^{13}\,  M_\odot$ to  $2.0\times10^{15}\,  M_\odot$.  
We refer to this set of clusters as the complete sample.
We  have
constructed their  evolutionary histories and, based  on their merging
histories,  
 we have classified  them into three categories according to
the mass ratio of the halos involved in the collision.

\begin{figure*}
\centering\includegraphics[width=16 cm]{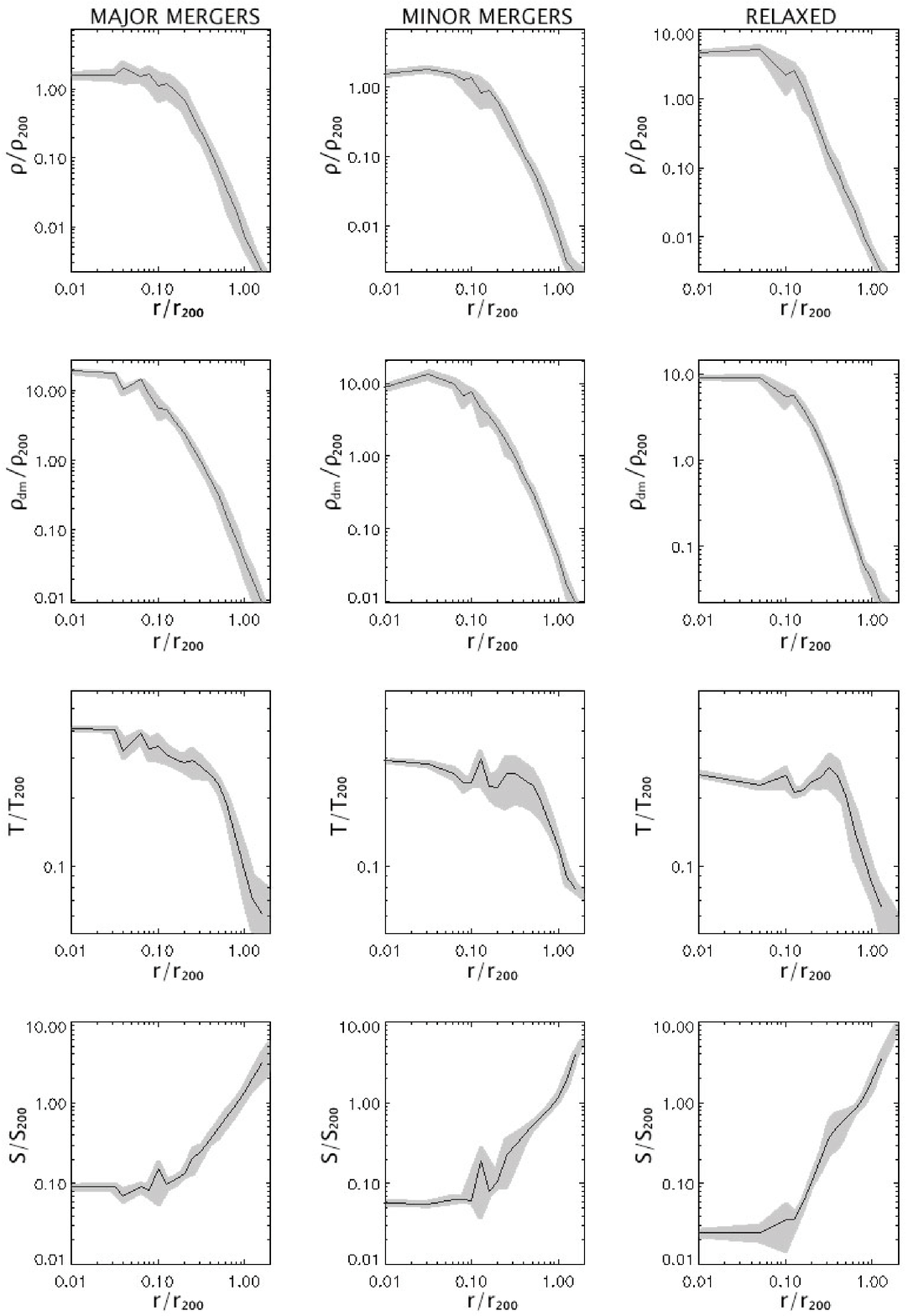}
\caption{Average radial profiles for all clusters belonging to 
each class: major mergers (left column), minor mergers (central
column), and relaxed (right column). From top to bottom, 
the first and second rows display gas ($\rho/\rho_{_{200}}$)
and dark matter ($\rho_{dm}/\rho_{_{200}}$) densities,
respectively, the third row shows mass-weighted temperature 
($T/T_{_{200}}$), and 
the fourth row represents entropy ($S/S_{_{200}}$). All profiles 
have been 
scaled by the mean value of the plotted quantities within the 
scale radius, $r_{_{200}}$. Continuous lines show the 
average for all the 
individual profiles of each class of clusters. Shadowed 
regions represent one $\sigma$ deviation.} 
\label{denrad1}
\end{figure*}

It  is  also convenient  to  adopt  a  timescale limit  since  mergers
occurring  at  a  very  early  epoch  would  not  have  any  important
consequence on the present properties of the clusters.  In this sense,
and only  for the  purpose of delimiting  the merger  events happening
recently,  we  have  defined  the  formation redshift  of  a  cluster,
$z_{for}$, as  the redshift at which  the cluster mass is  half of its
present virial  mass ~\citep{lacey93}. 
Thus,  we consider for  each cluster  only those
mergers that have relevant effects on its recent past.

Therefore, taking into account the formation redshift  
of the clusters, $z_{for}$, and the masses of the 
most (less)  
massive halo, $m_2\,\, (m_1)$, involved in the merger, we have
classified the clusters into  
three categories: 
\begin{itemize}
\item{} Major mergers. Those  systems where the mass ratio
$m_{2}:m_{1}$  is smaller  than $3:1$.   Therefore, a  major
merger  involves  clusters with  similar  masses.
\item{} Minor mergers. Those systems where the mass ratio is 
$3:1 < m_{2}:m_{1} < 10:1 $.
\item{} Relaxed halos. Those systems which have suffered mergers 
with very small halos, $10:1 < m_{2}:m_{1}$, or smooth accretion.
\end{itemize}

Out  of the  complete  sample, we  have  picked up  a subsample  which
contains the sixteen most massive galaxy clusters in the computational
box. They constitute what it would be referred to 
as the reduced sample,
and their main features are summarised in Table~\ref{tab1}.  Depending
on the particular analysis, that we will be interested in the following
sections  of  
this  paper,  we will  use  the complete  or the  reduced
sample,  respectively.  Concerning  their  merging classification, 
in the reduced sample we
have found five  relaxed clusters (R), three have  been categorised as
clusters with  minor mergers (MI),  and eight have been  classified as
major merger systems (MA).

Despite we have used an AMR code to carry out the simulation described
in  this paper,  due  to  numerical limitations,  a  biased sample  of
clusters -- with  a tendency to better describe  the most massive ones
-- has been produced.  These  artificial results could be overcome, in
future  applications,  by  performing  resimulations of  the  selected
clusters in the sample, although  this could prevent us from following
the mergers in a cosmological  context.  In any case, we consider that
the effects of  mergers would be more important  in those systems with
higher  masses --  well described  in  the present  simulation --  and
therefore, we believe that this bias has only minor consequences.

In  order to  analyse the  results of  our simulation,  we  will study
several  thermodynamical properties  which can  be  directly connected
with observational  data, and  which have been  widely studied  by all
sort  of different  simulations. In  addition to  the common  plots of
density, we  will also  study the behaviour of the  ICM temperature,
X-ray luminosity, entropy, and the internal and kinetic energies.

The ICM temperature will be defined as:
\begin{equation}
T =\frac{\sum_i T_i w_i}{\sum_i w_i},
\end{equation}
\noindent
where $T_i$ and $w_i$ are the temperature and the weight given to each
cell. In  most of  the applications in  the present paper,  the weight
will  be the  cell  mass, $w_i=m_i$,  and  therefore, this  will be  a
mass-weighted temperature.  In some particular cases, and for the sake
of comparison with observational data,  we will also use the so called
spectroscopic-like  temperature  ~\citep{mazzotta04}, $T_{sl}$,  where
the weight is $w_i=m_i \rho_i T_i^{-3/4}$ with $\rho_i$ the density at
the cell $i$.

A crucial  observable quantity, directly related with  the temperature
and  the density of  the gas, is the  bolometric X-ray  luminosity. In
simulations,  this quantity  can  be  computed by  adding  up all  the
contributions from each elemental volume of gas:
\begin{equation}
L_{_{X}}=\int_{V} n_{e} n_{i} \Lambda (T,Z) \, 
dV 
\end{equation}
\noindent 
where  $n_{e}$   and  $n_{i}$  are  the  electron   and  ion  density,
respectively,  and  $\Lambda$   is  the  normalised  cooling  function
depending   on  the   temperature   (T)  and   metallicity  (Z)   from
\cite{sudo93} (see Sec. 2.1).

The next thermodynamical quantity we  will pay a special attention is
the entropy, which is an extremely useful quantity providing a lot of
information~\citep{voit05}  about   the  evolutionary  state   of  the
clusters,  since it  records the  thermodynamical history  of  the ICM
produced  by the  gravitational and  non-gravitational  processes.  We
will adopt the following common definition for the entropy:
\begin{equation}
S=k_B T n_e^{-2/3} ,
\end{equation}
\noindent
where $n_e$ is the electron  number density and $k_B$ is the Boltzmann
constant.

Other  thermodynamical quantities  useful to  quantify the  effects of
mergers  and shocks as  a source  of feedback  are the  total internal
energy, $E_u$, and the total kinetic energy, $E_k$, which are given by
the following expressions: 
\be 
E_{u}=\int_{V_{vir}}\rho \epsilon \,dV
\en 
\be E_{k}= \frac{1}{2} \int_{V_{vir}}  \rho v^2 \,dV  
\en
\noindent
where $\rho$, $\epsilon$ and $v$  are the gas density, specific 
internal energy 
and the peculiar velocity of the gas fluid element, respectively. 

For the sake of comparison with observational data, 
it is also useful to define some characteristic quantities 
widely used in the literature. 
Similar to the definition of the virial quantities (see Sec. 2.2), 
 we introduce a characteristic radius, 
 $r_{_{\Delta}}$, such that the mean density enclosed within 
 this radius is $\Delta$ times $\rho_c$, 
and therefore, the mass is:
\be
\label{mdelta}
M_{\Delta}(<r_{_{\Delta}})=\frac{4}{3}\pi r_{_{\Delta}}^3 \Delta \rho_{c}.  
\en

Consequently, we 
follow the common definition for the  temperature,
\begin{equation}
\label{tvir}
T_{_{\Delta}}=\frac{GM_{_{\Delta}}}{2r_{_{\Delta}}}\frac{\mu m_p}{k_B},
\end{equation}
where 
$\mu$ and $m_p$ are the mean atomic weight and the proton mass. 
In the same manner, we define the entropy, 
\begin{equation}
\label{svir}
S_{_{\Delta}}=k_B T_{_{\Delta}} 
\left(\frac{\mu m_p}{f_b\rho_{_{\Delta}}}\right)^{2/3},
\end{equation}
with $f_b=\Omega_b/\Omega_m$.

In the following sections, and in order to compare with 
previous works, we will consider $\Delta=200$ or  $\Delta=500$,  
 depending on the particular case we compare with. 
All the quantities we have just described, are going to be used 
to analyse the results of the simulation.

\subsection{Merger history of selected clusters}

We have  constructed the merger  tree of the selected  galaxy clusters
tracking all the dark matter  particles that belong to a given cluster
backwards in time.  Figure~\ref{mertree} displays the merger trees
of three halos, which could be considered as prototypical ones of each
category  (i.e.,  relaxed,  minor  merger, and  major  merger).   The
clusters have  been selected  such that 
they have  very similar  masses and
sizes at  $z=0$.  The  merger trees  start at $z=0$  and plot  all the
parent halos  of the  final ones in  previous time steps  over several
output times of the simulation.

The total mass of each halo  is represented by a circle, whose size is
normalised  to  the  mass of  the  final  halo  at $z=0$.   The  lines
connecting  circles  between  different  times  inform  us  about  the
progenitors of a given halo at  a given time. In addition, the type of
line tells  us the amount of  mass transferred from  the progenitor to
the halo  at the  considered time.   Thus, a halo  at a  certain time
connected  with a  progenitor halo  at earlier  time by  a dash-dotted
line, means that up to 25\% of  its mass is due to the contribution of
that progenitor.  The same idea follows for other line types.  The aim
of this kind  of plot is to show the merger  history and the different
interconnections  over  time.   The  horizontal axis  is  designed  to
separate  halos  for plotting  purpose  only,  and  it has  no  direct
implication on  the position  of halos in  real space.   Vertical axes
show the redshift.

\begin{figure}
\centering\includegraphics[width=8 cm]{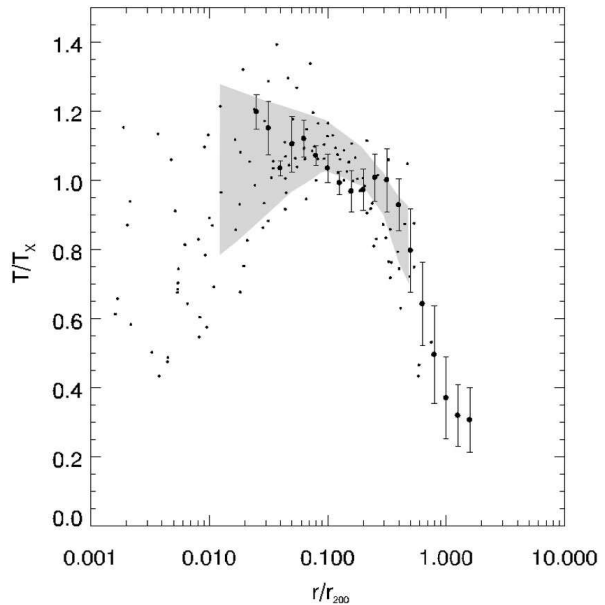}
\caption{Mean spectroscopic-like temperature profile for all the 
clusters in Table 1. Each cluster profile is normalised by its 
mean spectroscopic-like temperature ($T_X$) within $r_{_{200}}$. The 
normalised profiles are averaged in order to obtain the plotted 
mean profile. Dots represent the different bins equispaced in 
logarithmic scale with width 0.1 dex, and error bars are 1 $\sigma$ s.d.
The shaded region encloses the mean and 1 $\sigma$ s.d. 
temperature profiles from a representative sample of nearby 
clusters by ~\citet{pratt07}. The small black dots correspond to 
the temperature profiles of the clusters in the sample of 
~\citet{vik05}.} 
\label{pratt1}
\end{figure}

\begin{figure}
\centering\includegraphics[width=8 cm]{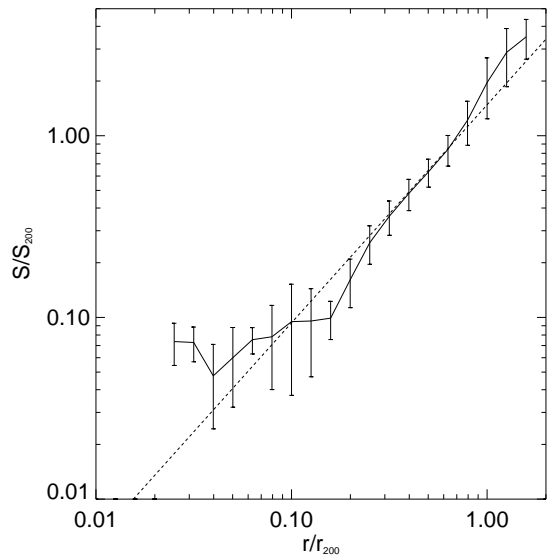}
\caption{Mean entropy profile for all clusters in Table 1
compared with the fitting in \citet{vhg05}.
The radial 
profiles for each cluster are obtained using equispaced 
logarithmic bins. All of them 
are normalised to their respective values of $S_{_{200}}$,  
and then, the mean profile is obtained. Continuous line 
represents the mean profile 
and error bars show 1 $\sigma$ s.d. Dashed line shows the fitting 
by \citet{vhg05}.}
\label{voit1}
\end{figure}

In  Figure~\ref{mertree} the  different  merger events  can be  easily
identified.  Whereas  the relaxed cluster  (right panel) has  a quiet
evolution, the  middle panel shows  a cluster suffering  three mergers
between  $z=0.79$  and  $z=0.65$.   By  comparing the  masses  of  the
different  halos  involved in  these  processes,  all  the events  are
classified as minor  mergers. In the left panel,  a cluster undergoing
several mergers at different times  is presented. Some of  the mergers
are minor ones, but there are major collisions at $z=0.99, z=0.79$ and
$z=0.72$.   In  order to  quantify  the  effect  of mergers,  we  will
correlate all these phases of  activity in the clusters evolution with
changes and effects on the different physical quantities.

It is  important to  notice that  some of the  merger trees  also show
halos that  break apart, that is,  lose mass and  reduce their sizes.
This  process operates at  two well  separated regimes  with different
causes. The  first group  is formed by  very small size  halos. These
halos are not really gravitationally bound and they can be easily
disrupted by  interactions with environment or with  other halos.  For
the halos  with larger masses, those  mass losses 
are  small, and they
are associated with tidal interactions.

\begin{figure}
\centering\includegraphics[width=8 cm]{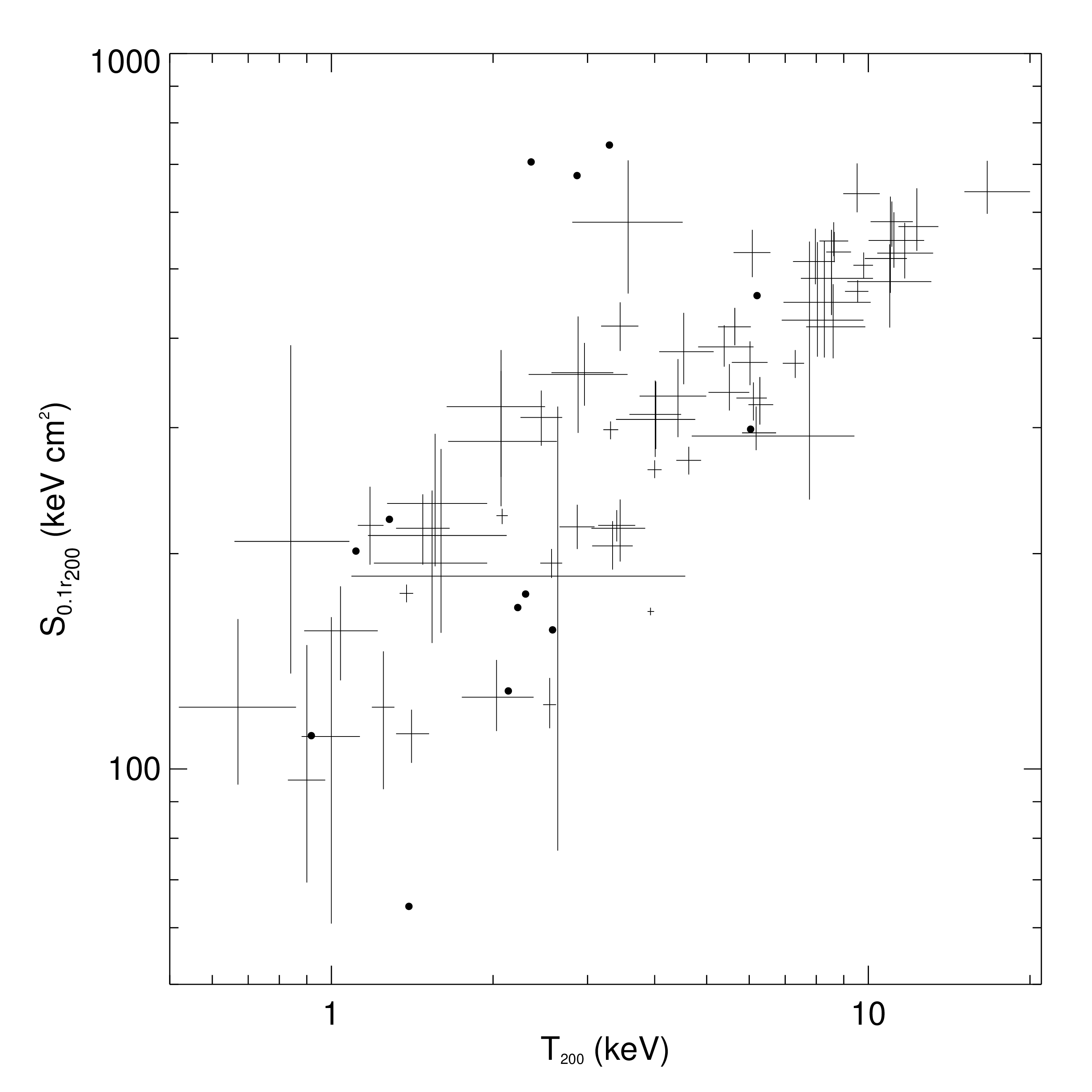}
\caption{Gas entropy at $0.1\, r_{_{200}}$ 
as a function of temperature, $T_{200}$.
Simulated clusters are represented by filled circles, whereas 
observational data with error bars stand for the 
~\citet{ponman03} sample. In order to compare simulations 
and observations, $T_{_{200}}\equiv T_X$ has been 
assumed.}
\label{ponmanS01}
\end{figure}

\subsection{Average radial profiles}
   
In  order to  analyse  the  main properties  of  the simulated  galaxy
clusters, we  compute radial  profiles for several  quantities.  These
profiles  are centred  at the  centre  of mass  of each  halo and  run
outwards  from the  centre  to  a distance  slightly  larger than  the
radius, $r_{200}$.  The bins  are equispaced in logarithmic scale with
widths  0.1 dex.   In all  the plots  displaying radial  profiles, the
radial coordinate is normalised to the $r_{200}$ at this time.

In Figure~\ref{denrad1}, we plot  averaged radial profiles for several
quantities for the clusters in  the reduced sample (see Table 1).  All
the profiles are scaled by the plotted quantities at $r_{200}$ defined
according  to  Eqs.(\ref{mdelta})-(\ref{svir}).   The mean  profiles
(continuous lines) are  computed by averaging all the  profiles of the
clusters  of each  class.  The  right  column stands  for the  relaxed
clusters, the central column represents the minor merger clusters, and
the  left column  displays  the major  merger  clusters.  The  plotted
quantities    are   gas    ($\rho/\rho_{200}$)    and   dark    matter
($\rho_{dm}/\rho_{200}$)  densities,  temperature  ($T/T_{200}$),  and
entropy  ($S/S_{200}$).  The  continuous  lines stand  for results  at
$z=0$ and shadowed regions mark one $\sigma$ deviation. 
Let us stress
that,  in Figure~\ref{denrad1}  and in  the following  ones  -- unless
explicitly  stated --,  we consider  mean profiles  rather  than median
profiles.

A detailed  analysis of Fig.~\ref{denrad1} shows the  main features of
the  three  categories  in  which  we have  classified  the  different
clusters. The comparison of  gas and dark-matter density profiles does
not  show  notable differences.   Whereas  for  the  gas density,  the
relaxed  clusters exhibit  a  slightly higher  density  at the  centre
compared with the  minor and major merger clusters,  the behaviour for
the dark  matter is  the opposite, having  the major
merger  clusters  a higher density.   In any  case,  the  profiles  are
consistent with  the expected characteristics of  density profiles for
galaxy clusters.

Concerning the temperature profiles, there are no dramatic differences
either.  All  clusters, in  the reduced sample,  show a  central core
with an almost constant  temperature and a declining profile outwards.
This   result   is   compatible   with   observational   data   (e.g.,
~\citet{degrandi02}),  and   with  the  idea  of   a  quite  universal
temperature  profile for  the galaxy  clusters  ~\citep{loken02}.  The
major  and minor  merger cluster  profiles have  very  similar central
temperatures,  although the isothermal  core is  larger for  the major
merger clusters.  The relaxed clusters  have a bigger  isothermal core
with a slightly lower value of the temperature compared with the major
and minor merger clusters.

\begin{figure*}
\centering\includegraphics[width=18 cm]{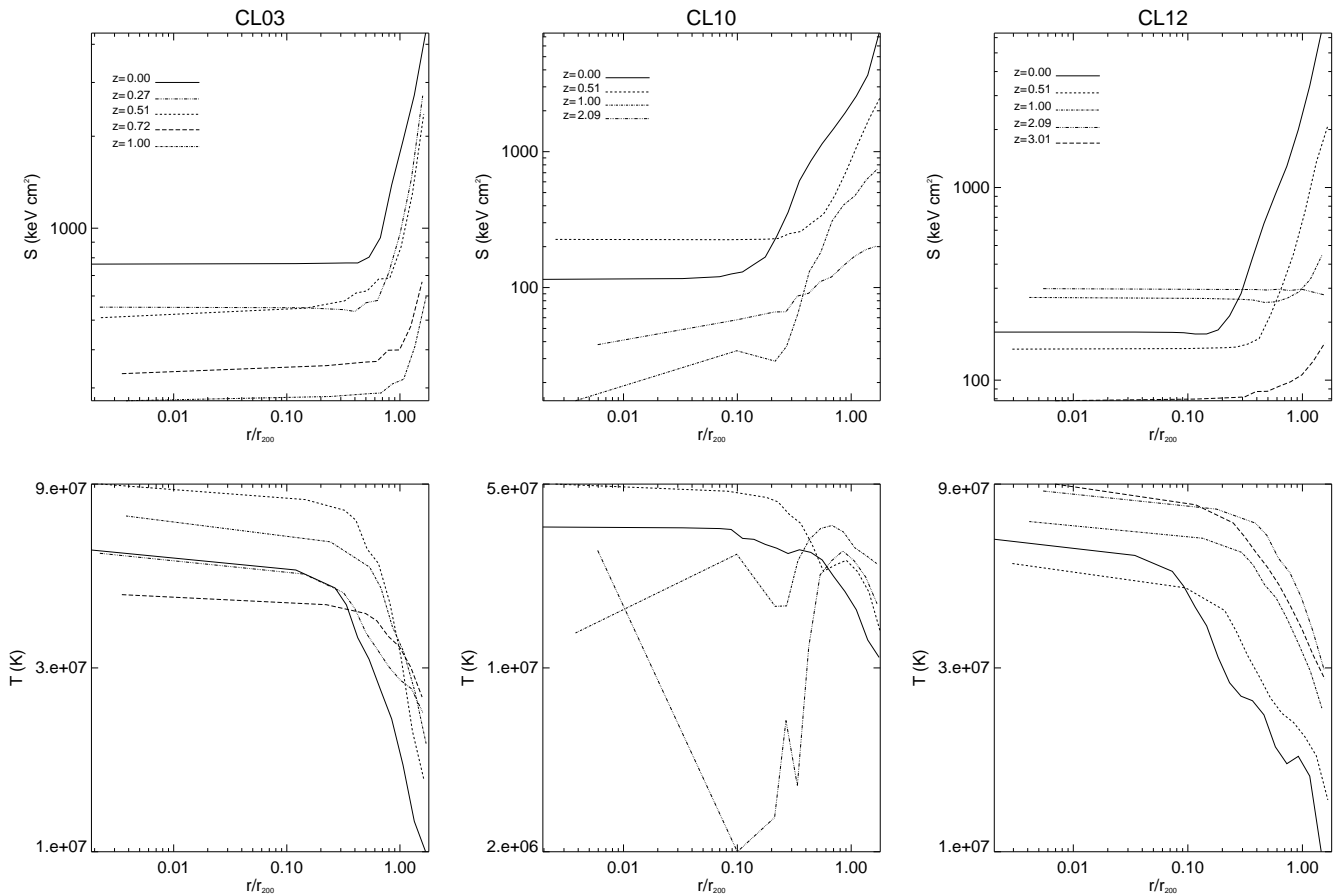}
\caption{Radial profiles of 
entropy ($S$) and temperature ($T$) at different redshifts. 
Left column shows the result for a cluster representing the 
major merger clusters, central column for a minor merger cluster and 
right column for a relaxed cluster.}
\label{st-shocks}
\end{figure*}

The temperature profiles  of the most massive clusters  in our reduced
sample  do not  exhibit a  drop in  the temperature  in  their central
regions. Apparently, this could seem to differ from the results of the
simulations by  ~\citet{kay07} or the observational  data presented by
~\citet{vik05} or  ~\citet{pratt07}. These last  observational results
show clusters  with temperature profiles with drops in  their central
temperatures.  This  effect  is   more  outstanding  in  the  case  of
~\citet{vik05}. The nature of this discrepancy amid both observational
results could be related with  the use of different instrumentation in
order to  obtain the  data of  both samples. It  must be  noticed that
whereas   ~\citet{pratt07}    used   XMM-Newton,   the    results   of
~\citet{vik05}  were obtained  using  CHANDRA --  with higher  angular
resolution.   This   could  explain  that  some  cool   cores  in  the
~\citet{pratt07}  sample were  not  properly resolved.   In order  to
compare with  the results  in ~\citet{pratt07} and  ~\citet{vik05}, we
have calculated the  spectroscopic-like radial temperature profile for
each cluster  of our reduced sample  (clusters listed in  Table 1) and
normalised them to their respective mean $T_{sl}$ within $r_{_{200}}$,
that we denote as $T_X$.   In Figure~\ref{pratt1}, we compare the mean
of  all these radial  temperature profiles,  represented by  dots with
error bars  (one $\sigma$ standard deviation),  with the observational
results  in ~\citet{pratt07}  marked  as the  shaded  region, and  the
results in ~\citet{vik05} represented  by small dots \footnote{It must
be  mentioned that  in  ~\citet{vik05}, the  temperature profiles  are
plotted  against   radial  coordinate  normalized  to  
$r_{_{180}}$. We  have ignored this small  correction without relevant
effects for the purpose of  the actual comparison.}.  Our results are
consistent with these  observational data in an  average sense,
being slightly more similar to the data of ~\citet{pratt07}.

We can understand our result, if  we keep in mind that the clusters we
are considering are  the most massive ones in our  sample.  As we will
discuss in  more detail  in Sec. 3.4,  we have  found that there  is a
strong anticorrelation between the  drop of temperature in the central
region and the mass of the cluster.  Therefore, the larger the mass of
the cluster the smaller the  number of clusters with central gradients
of temperature.  However,  if all the clusters in  the complete sample
are considered,  then a relevant  fraction of the  population ($\sim
16\%$) shows temperature profiles with central gradients (see Sec. 3.4
for more details).

More  interesting is  the analysis  of  the entropy  profiles. In  all
cases,  the   clusters  have   entropy  cores  and   profiles  outside
$0.2\,r_{200}$ compatible  with a power law  $S(r) \propto r^{\alpha}$
~\citep{tozzi01}.   In   previous  work   carried   out  by
~\citet{vhg05},  the authors performed  several non-radiative  SPH and
AMR simulations, and studied the main features of the entropy profiles
of the  galaxy clusters in  their numerical samples.   Besides several
differences  in  the  inner  cores,  all  clusters  in  their  sample,
regardless of the numerical technique used, showed very 
similar entropy
profiles outside  a region  around $0.2\,r_{200}$. In  particular, for
the AMR simulation,  they found that the entropy  profile in the outer
regions  can   be  better  fitted   by  the  power   law  $S(r)=1.43\,
S_{200}(r/r_{200})^{1.2}$. In Figure~\ref{voit1},  we compare the mean
entropy profile  of all the  clusters in Table  1 with the  fitting by
~\citet{vhg05}.   Continuous line  represents  our mean  entropy
profile with  $1\sigma$ error  bars.  Dashed line  stands for 
~\citet{vhg05} fitting. Our results seem to be compatible 
with the fitting
in the outer part of the profiles, whereas in the inner region, where
cooling  and feedback  processes  could be  relevant, differences  are
expected.

In  order to  compare our  results  with observational  data, we  have
looked at the values of the entropy  at $0.1\, r_{_{200}}$,  
$S_{_{0.1\,  r_{200}}}$,     
and   compared    them   with    previous    data   by
~\citet{ponman03}. In Figure~\ref{ponmanS01} we plot the observational
data (with  error bars) by ~\citet{ponman03} together  with the values
for the clusters  in our reduced sample with  temperatures higher that
$1\,  keV$  (filled   circles).   The  points  representing  simulated
clusters match well with the observational 
data apart from three clusters 
that are  marginally compatible. These three objects  turn out to have
some peculiarities  as they have suffered quite  recent merger events.
The values of the entropy at the very center of the simulated clusters
are  also similar to  recent Chandra  observations ~\citep{morandi07}.
Therefore,  our results  seem  to be  reasonably  compatible with  the
observations   taking  into  account   all  the   simplifications  and
limitations of our approach.

Coming back to  the comparison of the results  according to the merger
history of  the clusters,  the generic shape  of the  entropy profiles
does not depend systematically on the  mass or temperature of
the clusters,  in agreement with  observations ~\citep{ponman03}.  The
sizes  of  the cores  are  similar in  the  relaxed  and minor  merger
clusters and slightly larger in the major merger ones.  As it would be
naively expected, the  entropy floor in the relaxed  clusters is lower
than for  the minor merger clusters,  and this one is  also lower than
for the major  merger clusters.  Although the differences  seem not to
be dramatic,  they are  clearly visible in  the mean  profiles.  These
differences in the value of the  entropy in the core, would be a clear
consequence of the different evolutionary histories of each cluster.

\begin{figure*}
\centering\includegraphics[width=18cm]{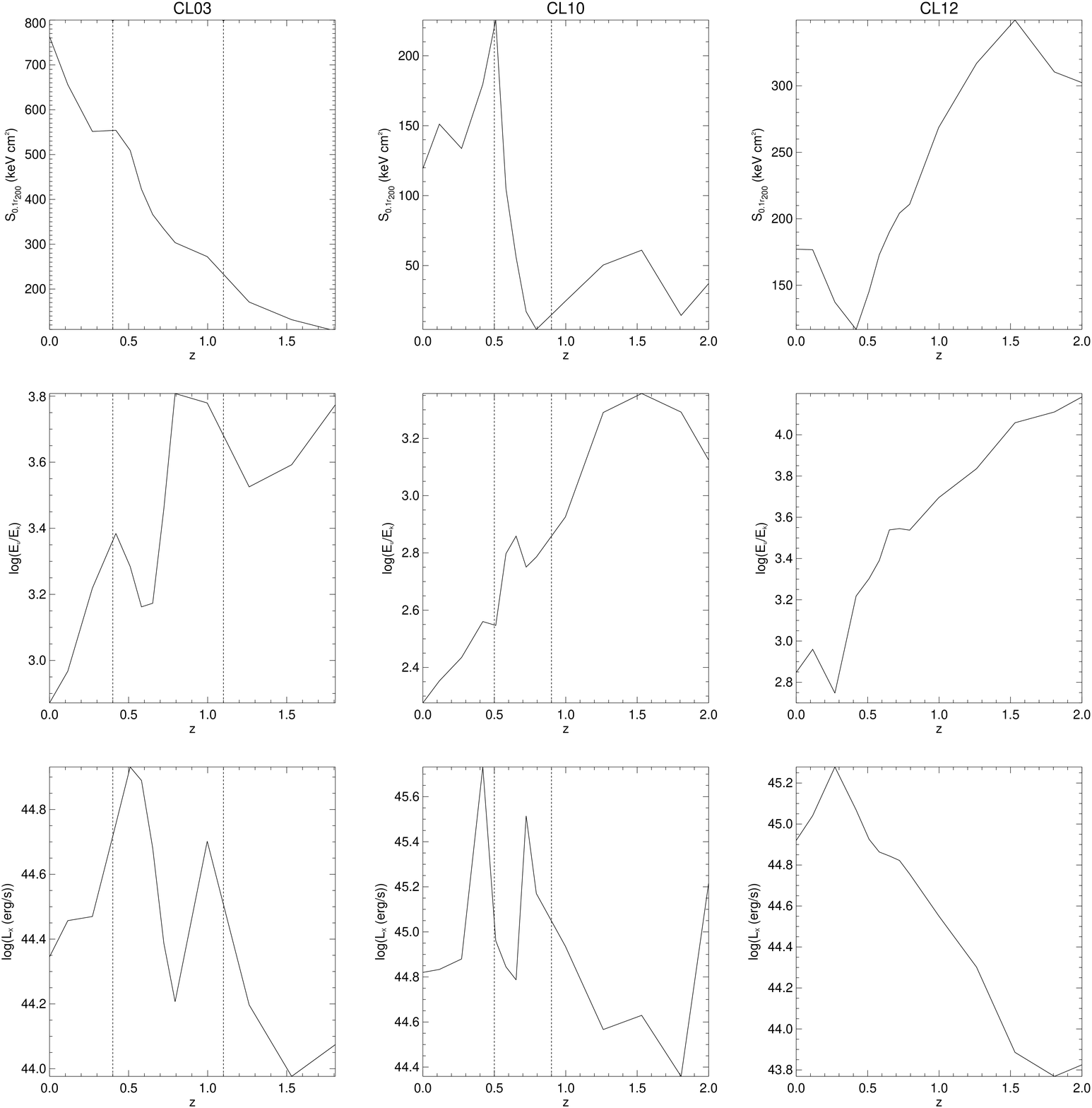}
\caption[]{Time evolution of the average  entropy ($S_{0.1r_{200}}$) 
within the inner 10\% of
the  radius $r_{_{200}}$,  the ratio
between total internal ($E_{_{U}}$) and kinetic ($E_{_{K}}$)  energies, and 
the X-ray  luminosity ($L_{_{X}}$), both 
evaluated within $r_{_{200}}$.  
Left  column  shows the  result  for  a  cluster
representing the  major merger clusters, central column  for a minor
merger cluster and right column for a relaxed cluster. 
Vertical lines delimit the time interval when mergers occur.}
\label{ste-shocks}
\end{figure*}

\subsection{Merger effects}

Merger events can produce shocks and compression waves in halos. Their
effects could be an efficient way to transfer part of the gravitational
energy associated to the collisions to the ICM of the final halo after
the mergers.  In   this picture, the    role  of turbulence and mixing
phenomena is  crucial as  a way to  redistribute this energy  into the
ICM.

The study of these scenarios  requires a numerical scheme able to
tackle with  an accurate description of shock  waves, strong gradients
as  well as  to describe  the turbulence  associated to  those violent
events. As it  has been discussed in Sec. 1,  the ability of different
numerical techniques to describe these  phenomena is still a matter of
debate. 

We focus in this Section on  the effects of different merger events in
the thermodynamical properties of the ICM.  In order to do so, 
we select the same  three  
clusters than  in  Sec. 3.1.   Each  one of  them
represents one of the three groups  of clusters, and as it was already
mentioned, they have been chosen in  such a way that they have similar
masses and sizes at $z=0$.

In  order to discuss  the effects  of different  merger events in the
thermodynamical    properties    of    the    ICM,    we    show    in
Figure~\ref{st-shocks}  the  radial  profiles  of  entropy  ($S$)  and
temperature  ($T$) at  several  redshifts for  the selected  clusters.
This figure  can be correlated  with Fig.~\ref{mertree} to  detect the
effect  produced  by  the  merger  events.   Lines  representing  high
redshifts must be taken carefully.  They correspond to early stages of
the  clusters  formation when  these  structures  are  far from  being
relaxed and, therefore, the radial profiles are not really meaningful.

The relaxed  cluster shows a  higher entropy and temperature  at high
redshifts, with a tendency to  reach a relaxed state around $z\sim0.5$
with  small changes.  The tendency  for  the minor  merger cluster  is
similar for the  lines displaying $z\sim 2$ and $z\sim  1$, that is, a
reduction of the  value of the entropy core.  However, associated with
the minor merger events, there  is a significant increase in the value
of the entropy  core which, eventually, ends up in  a reduction of the
temperature and  entropy at  $z\sim 0$ with  respect to the  values at
$z\sim 0.5$.  In the case  of the major merger cluster, between $z\sim
1$ and  $z\sim 0.6$  when the  major mergers take  place, there  is an
increase  in the  entropy  and  a reheating  as  the temperature  also
increases. Later on, the cluster cools to a lower temperature but part
of the energy of the merger has been released in the cluster which has
a higher entropy.  It must be  noticed that the values for the entropy
are  considerably larger than  for the  other two  clusters previously
discussed.

So  as  to  assess  more   clearly  the  effects  of  mergers  in  the
thermodynamical properties of  ICM, we show in Figure~\ref{ste-shocks}
the time evolution of  several quantities: the averaged entropy within
the 10\%  of $r_{_{200}}$ ($S_{0.1r_{200}}$),  the ratio  of the
total  internal  to kinetic  energies  ($E_{_{U}}/E_{_{K}}$),   
 and the integrated  X-ray luminosity ($L_X$), both  within the 
radius $r_{_{200}}$.  As in previous plots, each column 
represents the results of a
cluster representing one of the three classes.

For the relaxed cluster, the situation is simple. The early stages of
cluster  formation  have  left   high  entropy  and  internal  energy.
However,  with  the  time  evolution,  the cluster  cools  and  loses
internal  energy, creating a  lower entropy  core, and  increasing the
X-ray  luminosity.  The  minor  merger cluster  exhibits  a  different
history. As  the cluster forms a  bit later than the  previous one, it
begins  with a  lower entropy  and internal  energy compared  with the
relaxed cluster.  The time zone  when mergers happen --  delimited by
the vertical lines -- can  be clearly connected with important changes
in the cluster  evolution.  The first minor merger  boosts the entropy
level and  the internal energy,  indicating that some energy  has been
injected in  the system.  This energy  reheats the ICM  and produces a
decrease in  the luminosity by  delaying the cooling.  Later on,
the cooling takes over again dumping part of the energy, but leaving a
net increase in the entropy.   The history of the major merger cluster
is slightly different.  At the initial phase, the smooth accretion has
produced an increasing trend in  the core entropy, the internal energy
and  the  luminosity.   After   the  major  merger  the  situation  is
different.  Due to  the more  dramatic  effects of  the major  mergers
(higher  disruption, stronger  shock  waves, and  more turbulence  and
mixing) there is an increase in the entropy level, but associated with
an immediate loss  of energy due to radiation.   Another minor merger
produces some  minor changes but the  final state is  a cluster pretty
similar to the previous ones but with a significantly higher entropy.

\subsection{Cool cores and cluster mergers}

\begin{figure}
\centering\includegraphics[width=9 cm]{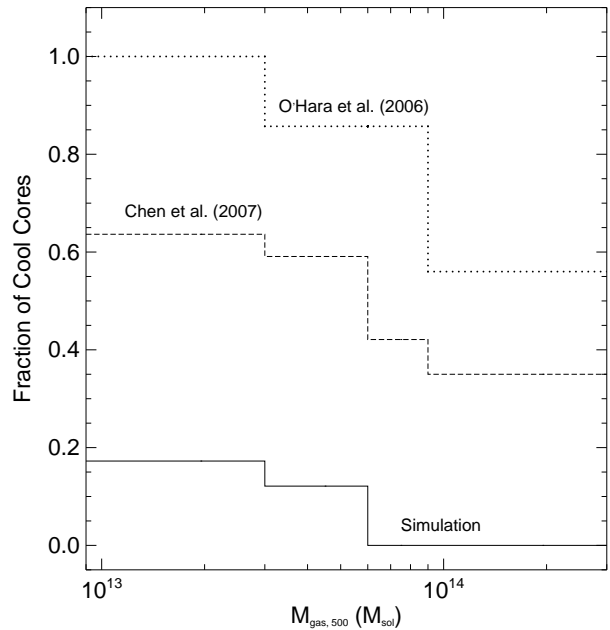}
\caption[]{Fraction of all galaxy clusters in our simulation 
that have CC vs the gaseous mass of the clusters. The simulated
clusters (continuous line) and the observed clusters from 
~\citet{chen07} (dashed line) and 
~\citet{ohara06} (dotted line) are binned in
five linearly equispaced bins.}
\label{ccc}
\end{figure}

It  is well  known  that  clusters of  galaxies  exhibit an  important
feature that allows 
to classify them into two separate populations, those
having cool cores (CC) and those others not having cool cores (NCC).

Recently,  ~\citet{chen07}  concluded   that,  roughly,  half  of  the
population observed in  a sample with more than  hundred clusters have
CCs. The explanation for this dichotomy  is not clear and it remains a
matter of debate.  Several authors have studied this  problem by means
of numerical simulations. Thus, ~\citet{kay07} overestimate the number
of CC clusters since almost all their 
clusters show the presence of CCs.
However, ~\citet{burns08} claimed to be the first authors producing a 
simulation with
CC  and NCC  clusters in  the  same numerical  volume, although  their
abundance at  $z\sim0$ of CC clusters,  $\sim 16\%$, seems 
to be lower
than   the  observed   fraction  by   ~\citet{chen07},   $\sim  46\%$.
Interestingly, the  results presented  by ~\citet{kay07} are  based on
SPH simulations,  whereas those of ~\citet{burns08}  are  
obtained using an
Eulerian  AMR code.   It is  likely that  feedback processes  could be
directly involved  in the survival of  CCs in clusters, but it is also
possible  that  mergers  could  play  an important  role  erasing  the
presence of CCs ~\citep{poole06,burns08}.  In this Section, we analyse
our simulation  paying special  attention to the  
presence of CCs, and their relative abundances.

\begin{figure}
\centering\includegraphics[width=9 cm]{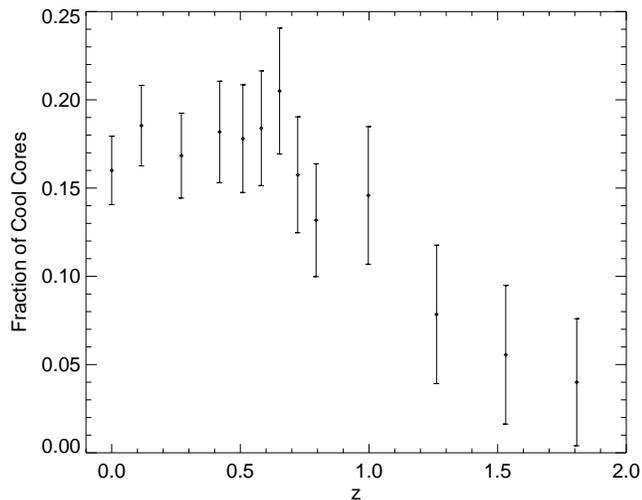}
\caption[]{Fraction of CCs as function of redshift for the 
simulated clusters in our complete sample. Error bars show $\sqrt N$
uncertainties due to the number counts.}
\label{ccz}
\end{figure}

Following ~\citet{burns08},  we define a  CC cluster as one with
 a $\geq  20\%$ reduction of its central  temperature compared with
the surrounding region. Using  this definition, we have classified all
the clusters in  our complete sample -- an  extended sample containing
all the clusters in the simulation -- into two groups: CCs and NCCs.

In Figure  ~\ref{ccc}, we plot  the fraction of  CCs as a  function of
gaseous mass at $z=0$. We have binned the clusters using five linearly
equispaced  bins in the  range $[10^{13},  10^{15}] \,  M_\odot$.  The
continuous line shows our results. For the sake of comparison, we have
used the samples of  ~\citet{ohara06} and ~\citet{chen07}, and we have
binned the clusters in these samples  using the same bins than for our
results.   The  dashed  and  dotted  lines correspond  to  Chen's  and
O'hara's results, respectively.

Our  results are extremely  similar to  those of  the ~\citet{burns08}
simulation,  where they  found  a  total fraction  of  $\sim 16\%$  CC
clusters (see figure  8 in this reference) and in  our case the number
is also  $\sim 16\%$.  As  in ~\citet{burns08}, our results  differ in
the   absolute  numbers   from  the   observational  data,   but  more
interestingly,  we have confirmed  the general  trend of  a decreasing
number of CC clusters with cluster mass.

Although, we have no clear explanation for the discrepancy between the
absolute number of CC clusters in our simulation and the observational
data by  ~\citet{chen07}, two plausible  explanations can be  given in
order to  interpret these results.  The first one  has to do  with the
fact  that  no metal-dependent cooling  has  been  considered  in  the
simulation. It  is known (see  for instance ~\citet{degrandi02,vik05})
that  some  clusters  can  show  strong  metallicity  gradients,  with
metallicities rising to solar  in the central regions. This limitation
of the  present simulation could produce some  artificial reduction of
the cooling,  specially in  systems where $kT  < 2\,  keV$. Therefore,
this shortcoming  could mimic, effectively, some  sort of uncontrolled
non-gravitational feedback.  The second possibility is  related with a
resolution  issue,  as no  resimulations  of  the  clusters have  been
performed. Therefore, despite  the use of an  AMR code, there could
exist some  resolution limitations.  This last possibility  seems much
less important, though.

We  have also  looked at  the  time evolution  of the  fraction of  CC
clusters. In  Figure ~\ref{ccz},  we plot the  fraction of CCs  in our
sample  as   a  function  of   the  redshift  from  $z\sim   2$  until
$z=0$. Again, our results are fully consistent with the simulations by
~\citet{burns08} and  show no important  change in the fraction of CCs
backwards in time, at least back  to $z\sim 1$. Our results are
in  contradiction with  observational evidences  showing  an important
variation in the fraction of CCs from $z=0.5$ ~\citep{vik06}.

Before $z\sim 1$, we find a  dramatic reduction in the fraction of CCs
with time.   As it would  be expected, the  abundance of CCs  would be
directly correlated  with the hierarchical formation  of the clusters.
At the epoch of cluster  formation, almost none of the clusters would
have a  CC. The  formation of CCs  would require the  establishment of
cooling flows which, eventually, and  through a slow process will form
the cool cores. However, once the clusters were fully formed, the major
mergers would destroy the CCs,  creating a population of NCC clusters.
It is clear that feedback processes  would also play a crucial role in
this  mechanism, but  in  the present  simulation,  where no  relevant
feedback mechanism  -- apart  of the gravitational  -- has  been taken
into account,  the effect of mergers  on the existence of  CCs is more
outstanding.  As  the mergers  are more dramatic  in the  more massive
systems, this  would explain the  anticorrelation of the  fraction 
of CCs and the mass of the clusters (see Fig.~\ref{ccc}).

In  order to  deepen  our knowledge of the  connection between  merger
activity  and the  presence of CCs in  clusters, we  have  studied the
dependence  of  the fraction  of  relaxed  clusters  with the  cluster
mass. If mergers  play an important role in the  existence of CCs, one
would expect  that the  systems that have  evolved quietly  (no recent
mergers) do have  cool cores. The comparison between  the fractions of
CC clusters and relaxed clusters is not direct as the establishment of
cooling flows and  the subsequent formation of CCs  could require long
time scales,  specially for the  smaller systems.  In 
any case,  and taking
the result with caution due to all the uncertainties, we have computed
the fraction of relaxed clusters which have central cores with cooling
times  shorter or  comparable to  the elapsed  time from  the clusters
formation   ($z_{for}$,   see   Sec.   3)  until   the   actual   time
($z=0$). Therefore,  these clusters  would have had  time to set  up a
CC. The result  shows the same trend than that in Fig.~\ref{ccc}, that
is,  the number  of relaxed  cluster (no  mergers) decreases  with the
cluster  mass.  This comparison  would show  that the  smaller systems
tend to  have a CC and  a quite evolution (no  merger events), whereas
the larger systems suffer the most important merger events and are NCC
systems.

\subsection{Scaling relations}

\begin{figure}
\centering\includegraphics[width=7.3 cm]{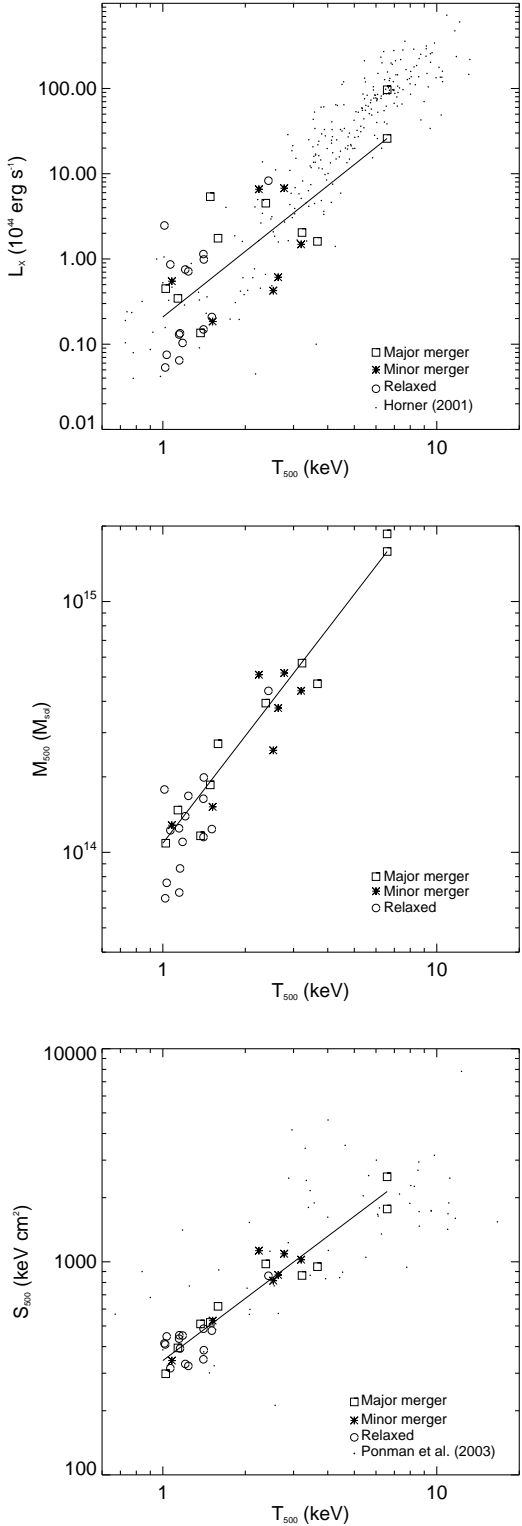}
\caption[]{Scaling relations  for our  galaxy cluster sample.  
The panels display, from top to bottom: integrated X-ray 
luminosity, mass, and  mean
entropy within the radius $r_{_{500}}$. In all the panels, the previous
quantities are plotted against the temperature, $T_{_{500}}$, 
computed according to Eq.~\ref{tvir}.
The different symbols represent the clusters in the 
sample,  
the continuous  lines  stand for the proper
fittings, and the small filled dots represent 
observational data by ~\citet{horner01} and ~\citet{ponman03},
respectively.}
\label{scale}
\end{figure}

The scaling relations are crucial  tools to study the galaxy clusters,
as  they connect  observables  like X-ray  luminosities, with  cluster
properties, namely, masses and temperatures.  Moreover, they can be an
excellent  way   to  check  the  behaviour  and   consistency  of  the
simulations, by comparing with the scaling relations obtained in other
simulations or  with observations.  

The  galaxy cluster  reduced sample  studied in  this paper  is biased
towards the  most massive clusters of our  simulation.  Therefore, the
statistical properties  of this sample  must be taken with  caution as
the  sample  is  far  from   being  complete,  due  to  the  numerical
limitations.  In the present  subsection, we have extended the reduced
sample  (see Table 1)  by considering  
all clusters, in the complete sample,  with temperature,
$T_{_{500}} \geq 1 \,\, keV$ (see Eq.~\ref{tvir}).

In  Figure~\ref{scale}, several scaling  relations are  plotted: X-ray
luminosity  (upper  panel),  mass  (middle panel),  and  mean  entropy
(bottom  panel)  within  the  radius  $r_{_{500}}$.  All  these  three
quantities  are  plotted against  the  temperature $T_{_{500}}$.   Our
results can be  fitted by the following scaling  relations: $L \propto
T^{2.5}$, $ M \propto T^{1.5}$,  and $S\propto T^{0.9}$.  In the three
relations,  we  have  plotted   all  the  clusters  with  temperatures
$T_{_{500}} \in[1.0,8.0] \, keV$.   This choice slightly increases the
number of  clusters of the original  sample presented in  Table 1. For
completeness, and in order to compare with observational data, we have
compared our scaling relations  with data by ~\citet{horner01} for the
$L-T$  relation  and  by  ~\citet{ponman03} for  the  $S-T$  relation,
respectively. These  data are displayed as  small dots in  the top and
bottom  panels  in Fig.~\ref{scale}.  The  results  for our  simulated
sample seem  to be consistent  with observational data,  leaving aside
all the uncertainties of such direct comparison.

Focusing on the effect of mergers, and for the sake of comparison with
previous works, let us assume that clusters that have had a relatively
quiet  evolution would  likely develop  a CC,  whereas  those clusters
involved  in  merger events  would  see  their  cool cores  distorted,
turning into NCC  clusters (see Sec. 3.4).  Under  this assumption, we
could consider  that the  galaxy clusters in  our sample,  labelled as
major and  minor mergers, could be identified,  broadly speaking, with
NCC  clusters   like  the   ones  studied  by   ~\citet{poole08},  and
analogously, our  relaxed clusters  would be the  CC clusters  in this
reference.  The results displayed in Fig.~\ref{scale} show some degree
of  segregation, with  most of  the  major and  minor merger  clusters
located at well separated regions  on the scaling relation plots.  The
minor merger  clusters sit, preferentially, at  an intermediate region
between  the major  merger clusters  and the  relaxed clusters.   As a
gross trend, the majority of clusters which have suffered mergers, are
placed in  zones with higher temperature and  higher luminosity, mass,
and entropy, respectively.

\begin{figure}
\centering\includegraphics[width=9 cm]{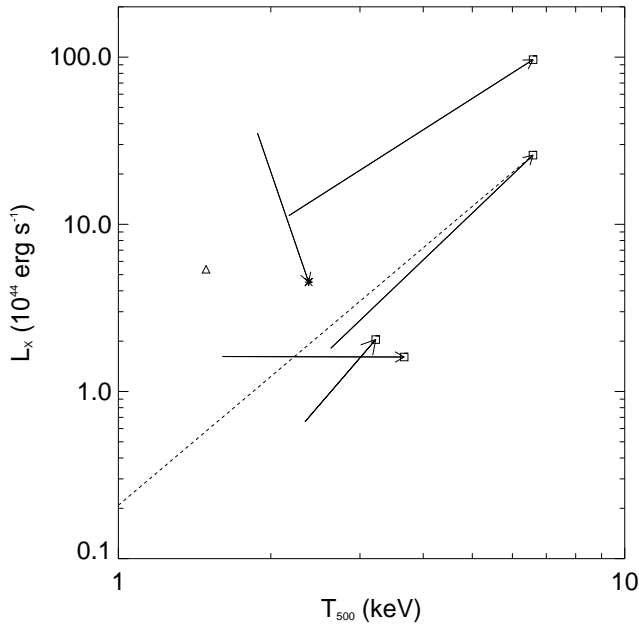}
\caption[]{The X-ray luminosity vs the temperature, $T_{_{500}}$, 
within radius, $r_{_{500}}$.
The plot shows, as vectors, the overall drifts 
 experienced 
by the most massive major merger clusters  in their 
positions at the L-T relation from  
$z=1.5$ until $z=0$. The clusters, at $z=0$, are 
labelled by squares, except two particular cases, identified
by a triangle and a star, discussed in the text. 
Dashed line stands for the fitting of the L-T 
relation for the whole sample.} 
\label{scale2}
\end{figure}

Mergers, specially the major  ones, typically boost clusters along the
L-T    relation   but    not   parallel    to   this    relation.   In
Figure~\ref{scale2},  we plot  -- for  the major  merger  clusters in
Table  1 --  the overall  drifts  experienced by  these clusters  from
$z=1.5$ until $z=0$. This evolution is illustrated by vectors starting
(finishing) at  the values for  L and T  at $z=1.5$ ($z=0$).   All the
vectors could be decomposed  in two components representing the change
in  temperature  and  luminosity.   Two  special  clusters  deserve  a
particular discussion.  The first one, represented by  a triangle, has
no arrow  associated.  This is due  to the fact that  this cluster has
been classified  as major merger  but it marginally satisfies  the 1:3
condition  in the  mass  ratio.  Therefore, it  is  a transition  case
between  major and minor  clusters, according to our  definition of
merger.  However, as  we  will discuss  later,  consistently with  its
evolution in the L-T plane, it behaves like the minor merger clusters.
The second particular  case, represented in the plot  by a star, shows
the evolution  of a  cluster with a  extremely strong cooling  flow at
$z=1.5$. A careful study of  the radial temperature of this cluster at
that  time,  shows  an   extremely  relevant CC.  During  the
evolution, the  merger event  substantially reduces the  cooling flow,
and disturbs  the CC,  although the cluster  remains radiating.
Interestingly,  all  clusters,  except  the  one  represented  by  the
triangle, show  a net increase in  temperature. Concerning luminosity,
letting aside  the two  particular cases just  mentioned, the  rest of
clusters shows an increase in luminosity.  We have performed a similar
analysis for the minor merger clusters, and they do not show any clear
trend and seem to behave very similar to the relaxed clusters.

There  is  a  clear  bias  in  the treatment  of  small  clusters  and
groups. Nevertheless the results of our simulation are consistent with
previous  results,   specially  considering  that  it   has not  been
introduced any  other pre-heating  or feedback mechanism,  besides the
one from the star formation  (very poor in the present simulation) and
the so-called gravitational feedback  (shock waves, mergers, etc).  In
any case, even though the sample  can be limited, we wish to stress that
the individual  properties of  each of the  most massive  clusters are
well defined.

\section{Discussion and conclusions}

We have presented the results of a hydro and dark matter simulation of
a  moderate  size  volume  of  the  Universe in  the  framework  of  a
concordance cosmological model.   The simulation follows the evolution
of gaseous and dark matter components.  Other relevant processes, like
heating and cooling for a primordial gas, have been 
also taken into account.

The  main idea of  the present  paper has  been to  study the  role of
galaxy cluster mergers as source  of feedback of the ICM.  The general
picture  of  our simulation  shows  how  mergers  stirred the  ICM  by
producing shocks and sound waves in scales comparable to the dimension
of  the cluster.  These phenomena  produce turbulence  in the  form of
large eddies characterised by Reynolds  numbers, $Re \sim 10^3$, which
are  values  accessible  to   present-day  numerical  simulations
~\citep{ricker2001}.

Recent  work  by  ~\citet{agertz07}   has  shown,  in  some  idealised
situations, how  the different  numerical techniques, namely  SPH and
AMR,  produce  different results  when  describing  the formation  and
evolution of  eddies (Kelvin-Helmholtz) instabilities.  According to
this last  reference, there are no substantial  differences in results
obtained with  both techniques if  the evolution time is  smaller than
the   characteristic   time    of   formation   for   Kelvin-Helmholtz
instabilities,  $\tau_{KH}$. This  characteristic time  can  be 
estimated for the case of  a cluster merger assuming similar densities
for the clusters,  a characteristic length scale of  a few hundred  
of
$kpc$, and a relative velocity of the order of 
one thousand km/s; with all
this  conditions,  $\tau_{KH}$ turns  to  be of  the  order  of a  few
hundreds of millions of years.

In the  particular case  of galaxy cluster  mergers, the  typical time
scales  of  evolution  after  a  merger  are  much  larger  than  this
$\tau_{KH}$,  and therefore, it  is likely  that 
the  particular numerical scheme
used  to  simulate   such  scenario  could  play  some   role  in  the
results. This has motivated us to study 
this problem by means of an AMR
technique,  and  as  a  complementary  work  to  the  studies  already
published.   Therefore, we have  used an  AMR Eulerian  code specially
designed for cosmological applications, with excellent capabilities to
deal  with strong  gradients, shock  waves, and  low  density regions.
These  kind  of  codes   also  describe  properly  some  instabilities
phenomena  such as  Kelvin-Helmholtz, Rayleigh-Taylor and, in general,
turbulence and mixing processes (see 
the recent work by ~\citet{mitchell09}).

In order  to study the effect  of mergers as feedback  source, we have
extracted and followed the evolution  of the galaxy cluster like halos
in our  simulation. These  halos have been  studied directly  from the
simulation  and without any  resimulating scheme.  This has  implied a
limitation due  to numerical restrictions. Since  the numerical scheme
has the tendency to better  resolve the most massive halos, our sample
is biased towards these systems. However, it is in those large systems
where  the effects of  mergers would  be more  relevant. On  the other
hand, this apparently  drawback of lack of resolution  on small halos,
is compensated  by
the  advantage that mergers  can be followed  in a
consistent  way  as they  naturally  occur  in  the evolution  of  the
simulated volume of the  Universe. Thus, no controlled collisions have
to be imposed a priori.

We have assumed a definition of galaxy cluster merger depending on the
mass  ratio  of  the  halos  involved in  the  merger.  As  additional
condition, we  have introduced  a time limitation, in such a  way that
only  mergers occurring  in the  recent past  are taken  into account.
Thus, if  masses are similar -- between  1 and 1/3 --  we define those
events as major mergers. Events  with larger ratios in masses are
classified  as  minor  mergers.   Finally, clusters  evolving  without
relevant merger events are designed  as relaxed.  Once we have assumed
this criterion  
to group the clusters in our simulation, we have studied
the  radial profiles  of  each cluster  for  the gas  and dark  matter
densities,  temperature and  entropy.  In  order to  compare  the main
differences among the three classes, we have computed average profiles
for each group. The forms  of the different profiles are basically the
same for  the three categories,  indicating no substantial  changes in
the  physics  of   clusters.   However,  there  is  a   trend  in  the
normalisation.   The relaxed  and minor  merger clusters  have similar
values  of  all quantities,  whereas  the major  merger  clusters  are
slightly hotter and with higher entropy.

The previously mentioned trends in entropy and temperature of clusters
depending on their evolutionary history, can be quantify by looking at
a  representative  cluster  of  each  class  and  comparing  the  time
evolution of their profiles of temperature and entropy.

In the same manner, the effects associated with cluster mergers can be
traced in the time evolution  of global quantities like the entropy in
the inner 10\% of the virial  radius, the ratio of internal to kinetic
energy, or the X-ray luminosity.

In all cases,  cluster mergers release energy which  ends up partially
in the final object.  The amount of energy locked in the final cluster
is significantly  larger for cluster mergers of  similar masses (major
mergers).

We have  also considered  all the clusters  in the  simulation without
differentiating amid the merging  activities.  These results have been
compared       with        previous          simulations
~\citep{vhg05,burns08,kay07}     and    with     observational    data
~\citep{ponman03,vik05,pratt07},  paying   special  attention  to  the
entropy and  temperature profiles. Our  results seem to  be consistent
with both, simulations and observations, in an average sense. However,
there are still  important differences like the lack  of central drops
in the temperature profiles of the most massive clusters.

The fraction  of clusters in our  sample that has cool  cores has been
computed  at  several  redshifts.  At  $z=0$, our  results  are  fully
compatible with previous AMR simulations by ~\citet{burns08}, although
seem  to   differ  with   the  results  of   the  SPH   simulation  by
~\citet{kay07}. We  have compared  the fraction of  cool cores  in our
simulation with  the observational  data by ~\citet{chen07}  showing a
similar  trend,  that is,  the  number  of  clusters with  cool  cores
decreases with the cluster mass.

Given the  fact that in  our simulation the gravitational  feedback is
the relevant feedback mechanism, we have tried to correlate the merger
events with the  existence of cool cores.  In order to  do so, we have
computed the  fraction of relaxed  clusters (no merger activity)  as a
function of the cluster  mass. Interestingly, the fraction of clusters
with cool cores,  and the fraction of relaxed  clusters as function of
the  cluster mass,  show  a very  similar  trend.  Unfortunately,  the
absolute numbers  of CCs  in our simulation  and the  observations are
quite  different.   We  suggest  two   possibilities  explaining  this
discrepancy.  The first one  would be  related with  the fact  that no
metal-dependent cooling  has been considered in  the simulation.  This
simplification could  make the cooling more  inefficient, specially at
the central regions of the clusters. The second reason would be linked
with a  lack of resolution, which  appears to be  quite unlikely given
the actual  features of  the considered simulation.   In any  case, it
seems clear  that there is an  evident link between  the merger events
and the no existence of cool cores.

On the  other hand, the time  evolution of the fraction  of cool cores
shows that this quantity has  not changed substantially from $z\sim 0$
to  $z\sim 1$.  This result  is compatible  with  previous simulations
~\citep{burns08}   but  in   disagreement   with  observational   data
~\citep{vik06}.

The cluster sample  analysed in this paper is limited  due to the fact
that no resimulations  have been done, and therefore,  although an AMR
code  has   been  used,   there  are  still   resolution  limitations.
Nevertheless, we have analysed  the scaling relations derived from our
sample. Our results for $L \propto T^{2.5}$, $ M \propto T^{1.5}$, and
$S\propto T^{0.9}$  are consistent with  previous results that  do not
introduce any extra reheating or feedback.

We  have found  some degree  of segregation  in the  scaling relations
depending on whether the clusters  have or have not undergone a recent
merger.  The  systems  that have experienced merger events 
are  usually located  at  high  temperatures,
luminosities,  masses, and entropies,  respectively, at  the different
scaling relation plots.  These results could be comparable with recent
works  looking at  the  existence or  not  of cool  cores in  clusters
~\citep{burns08,poole08}. The analysis of  the time evolution of major
merger clusters  in the  L-T relation, has  shown that  these clusters
have a tendency  to move towards regions of  this relation with higher
temperature and luminosity.  This tendency is similar to that found by
~\citet{hartley08}, where  authors investigate  the L-T relation  in a
large simulation with a strong preheating.

A clear improvement for future work would be to increase the number of
clusters  in  the sample  by  simulating  larger  volumes with  higher
resolution.   Therefore, it would  be feasible  to reliably  study the
scaling relations for each one  of the three families of clusters that
we have considered  in this paper.  In any case,  even when the sample
can be limited, the individual  properties of each of the most massive
clusters are well defined.

The role  of mergers as source  of feedback, transferring  part of the
gravitational  energy to  the thermal  energy,  is still  a matter  of
debate and  study.  Mergers are  crucial to understand  galaxy cluster
formation and  galaxy formation  scenarios as they  influence directly
the ICM properties.  Simulations  with higher resolution and including
more physical processes are needed in order to keep on quantifying the
role   of  mergers   in   the  hierarchical   scenario  of   structure
formation. In  parallel, some results  from simulations, like the ones
presented here,  can be considered in semi-analytical  models in order
to improve their description of the gas component.

\section*{Acknowledgements}
This work has  been supported by {\it Spanish  Ministerio de Ciencia e
Innovaci\'on}     (MICINN)     (grants    AYA2007-67752-C03-02     and
CONSOLIDER2007-00050).   SP thanks to  the MICINN  for a  FPU doctoral
fellowship. The  authors wish  to thank to  the anonymous  referee for
his/her      valuable     and     constructive      criticism,     and
J.M$^{\underline{\mbox{a}}}$.   Ib\'a\~nez and J.A. Font 
for useful  discussions and
comments. A. Vikhlinin kindly provided us with some of the observational 
data used to compare with our results.   Simulations  were   carried  
out   in  the   {\it  Servei
d'Inform\'atica de  la Universitat de Val\`encia} and  the {\it Centre
de Supercomputaci\'oó de Catalunya (CESCA)}.

\section*{APPENDIX A: Star formation}
 
The star formation  has been introduced in the  MASCLET code following
the ideas of \citet{yepes97} and \citet{springe03}.  In our particular
implementation, we assume that cold  gas in a cell is transformed into
star particles on  a characteristic time scale $t_*$  according to the
following expression: 
\be
\frac{d\rho_*}{dt}=-\frac{d\rho}{dt}=\frac{\rho}{t_*(\rho)}-
\beta\frac{\rho}{t_*(\rho)}= (1-\beta)\frac{\rho}{t_*(\rho)} \en
\noindent
where   $\rho$  and  $\rho_*$   are  the   gas  and   star  densities,
respectively.  The  parameter $\beta$ stands for the  mass fraction of
massive  stars  ($>8\,M_{\odot}$)  that  explode  as  supernovae,  and
therefore return to  the gas component in the  cells.  We have adopted
$\beta=0.1$,  a  value  compatible   with  a  Salpeter  IMF.  For  the
characteristic  star formation  time,  we make  the common  assumption
$t_*(\rho)=t^*_o(\rho/\rho_{_{th}})^{-1/2}$,       equivalent       to
$\dot\rho_*=\rho^{1.5}/t^*_o$  \citep{keni98}.  In  this way,  we have
introduced a dependence on the local dynamical time of the gas and two
parameters , the density threshold for star formation ($\rho_{_{th}}$)
and  the corresponding  characteristic time  scale ($t^*_0$).   In our
simulations     we     have      taken     $t^*_0=2\,     Gyr$     and
$\rho_{_{th}}=2\times10^{-25}\,g\,cm^{-3}$.  From  
the energetic point of
view,  we consider  that each  supernova  dumps in  the original  cell
$10^{51}\, erg$ of thermal energy.

In the  practical implementation, we have assumed  that star formation
occurs once every global time  step, $\Delta t_{l=0}$ and, only in the
cells at the highest level of refinement ($l=7$).
Those cells  at this level  of refinement, where the  gas temperature
drops below  $T <  2\times10^4\, K$,  and the gas  density is  $\rho >
\rho_{_{th}}=2\times10^{-25}\,  g\,  cm^{-3}$,  are suitable  to  form
stars.   In  these  cells,  collisionless  star  particles  with  mass
$m_*=\dot\rho_*\Delta  t_{l=0}\Delta x_l^3$ are  formed.  In  order to
avoid sudden changes in the  gas density, an extra condition restricts
the    mass    of    the    star    particles    to    be    $m_*={\rm
min}(m_*,\frac{2}{3}m_{gas})$, where  $m_{gas}$ is the  total gas mass
in the considered cell.

\end{document}